\documentclass[twocolumn,showpacs,amsmath,nofootinbib]{revtex4-1}

\usepackage{float} 
\usepackage{amsmath}
\usepackage{amsfonts}
\usepackage{amssymb}
\usepackage{amsthm}
\usepackage{graphicx}
\usepackage[colorlinks,citecolor=blue]{hyperref}
\usepackage{color}
\usepackage{flafter}
\usepackage[normalem]{ulem}

\usepackage[toc,page]{appendix}
\usepackage{makecell}
\usepackage{slashed}
\usepackage{tikz-cd}
\usepackage{makecell}
\usepackage{subcaption}
\usepackage{graphicx}

\allowdisplaybreaks

\definecolor{OliveGreen}{rgb}{0,0.6,0}
\definecolor{Orange}{rgb}{1.00, 0.65, 0}
\definecolor{Grey}{rgb}{0.43, 0.5, 0.5}

\newcommand{\Fig}[1]{Fig.~\ref{#1}}
\newcommand{\Eq}[1]{Eq.(\ref{#1})}
\newcommand{\nn}{\nonumber\\}

\newcommand{\<}{\langle}
\renewcommand{\>}{\rangle}
\newcommand{\ket}[1]{|{#1}\rangle}
\newcommand{\bra}[1]{\langle{#1}|}
\newcommand{\be}{\begin{eqnarray}}
\newcommand{\ee}{\end{eqnarray}}
\newcommand{\bpm}{\begin{pmatrix}}
\newcommand{\epm}{\end{pmatrix}}

\newcommand{\Tr}{{\rm Tr}}

\renewcommand{\Re}{{\rm Re}}

\newcommand{\ra}{\rightarrow}

\renewcommand{\v}[1]{{\boldsymbol{#1}}}

\newcommand{\s}{\sigma}
\renewcommand{\t}{\tau}

\newcommand{\G}{\Gamma}

\newcommand{\comment}[1]{}

\begin{document}

\title{Fractional Vortices, $\mathbb{Z}_2$ Gauge Theory, and the Confinement-Deconfinement Transition}
\author{Zhi-Qiang Gao$^{1}$}
\author{Yen-Ta Huang$^{1}$}
\author{Dung-Hai Lee$^{1,2}$}\email{Corresponding author: dunghai@berkeley.edu}

\affiliation{
	$^1$ Department of Physics, University of California, Berkeley, CA 94720, USA.\\
	$^2$ Materials Sciences Division, Lawrence Berkeley National Laboratory, Berkeley, CA 94720, USA.
}

\begin{abstract}
In this paper we discuss the classical 3D XY model whose nearest-neighbor interaction is a mixture of $\cos(\theta_i-\theta_j)$ (ferromagnetic) and $\cos2(\theta_i-\theta_j)$ (nematic). This model is dual to a theory with integer and half-integer vortices. While both types of vortices interact with a non-compact U(1) gauge field, the half-integer vortices interact with an extra interaction mediated by a $\mathbb{Z}_2$ gauge field. We shall discuss the confinement-deconfinement transition of the half-integer vortices, the Wilson and the `t Hooft loops and their mutual statistics in path integral language. In addition, we shall present a quantum version of the classical model which exhibits these physics.
\end{abstract}
\maketitle

\newpage

\section{Introduction}
More than three decades ago Lee and Grinstein \cite{Lee:1985aa} considered a modified 2D classical XY model described by the following Hamiltonian
\be
&&H=-J_1\sum_{\<ij\>}\cos(\theta_i-\theta_j)-J_2\sum_{\<ij\>}\cos2(\theta_i-\theta_j).\label{Hs}\ee
Here $\<ij\>$ denotes nearest neighbors. For $4J_2>J_1$, the angle difference $\theta_i-\theta_j=\pi$ is a local metastable minimum. Because of that, half-integer vortices around which the spin angles wind by $\pm \pi$ are local energy minimum configurations. Two half-integer vortices are connected by a string across which the nearest  neighbor angle difference is $\pi$.
The free energy difference per unit length between $\theta_i-\theta_j=2n\pi$ and $\theta_i-\theta_j=(2n+1)\pi$ is a function of temperature. At low temperatures the string tension is non-zero and the half-integer vortices are confined by a linear potential. At high temperatures, the configurational entropy balances the energy cost hence the string tension vanishes. Under such condition the half-integer vortices go through a confinement-deconfinement transition governed by the 2D Ising model. The resulting phases were discussed in Ref.\cite{Lee:1985aa}. In fact, the question concerning how the phases of Ref.\cite{Lee:1985aa} meet, and the universality class of the phase transition, are still actively studied today.

The purpose of this paper is to generalize the Lee-Grinstein model to 3D. Here the point-like vortices become vortex loops; the string between the half-integer vortices becomes a surface sustained by the half-integer vortex loops; the logarithmic vortex interaction becomes the interaction between the vortex loops mediated by a non-compact U(1) gauge field; the confinement-deconfinement transition of half-integer vortices is determined by whether the surface tension is finite. In Sec. II we mathematically generalize the standard duality transformation of the ordinary 3D XY model\cite{Dasgupta:1981aa,Fisher:1989aa} with only the $J_1$ term to show that the half-integer vortices couple to a $\mathbb{Z}_2$ gauge field, and the deconfinement transition discussed above is governed by the deconfinment transition in the 3D $\mathbb{Z}_2$ gauge theory\cite{wegner1971,Kogut:1979aa}.  In Sec. III we discuss the Wilson and `t Hooft loops \cite{Wilson:1974aa,thooft1979} and their linking phase, i.e., the anyon mutual statistics, in the deconfined phase. In Sec. IV we discuss the confinement-deconfinement transition using a quantum version of \Eq{Hs}. Conclusions are drawn in Sec. V and details of derivations are given in Appendix. The main results of this paper are (1) in the model governed by \Eq{Hs} where there is no built-in gauge fields, nonethless a $\mathbb{Z}_2$ gauge structure emerges. Therefore, the $\mathbb{Z}_2$ gauge field in our work  is entirely emergent. (2) In the deconfined phase of this $\mathbb{Z}_2$ gauge theory, a $\mathbb{Z}_2$ topological ordered phase coexists with conventional symmetry-breaking order. (3) A simple generalization of \Eq{Hs}, namely \Eq{znXY}, has emergent $\mathbb{Z}_n$ gauge structure and realizes the associated $\mathbb{Z}_n$ topological order. Last, but not least, all of the above results are obtained analytically.

Experimentally, the model in \Eq{Hs} can be realized in cold bosonic atomic gases interacting
via a Feshbach resonance \cite{Radzihovsky2008} in either 2D or 3D. In addition, based on a suggestion made in Ref.\cite{Fu2010}, recently it is suggested that an one dimensional chain of superconducting islands each harboring two Majorana zero modes\cite{Roy2020} actually simulates the model proposed in  Ref.\cite{Lee:1985aa}. Moreover, generalizing Ref.\cite{Fu2010} to 2D, several interesting works focus on two-dimensional Josephson arrays with four Majorana zero modes per superconducting island (the so-called Majorana toric code (MTC)) find similar phases and the same $\mathbb{Z}_2$ topological order\cite{Xu2010,Terhal2012,Roy2017} as in our work. However, because (1) the Hilbert space difference, and (2) the difference in the path integral action,  it is unclear to us what is the relation between our model and the MTC. However, it is clear that understanding such relation is an interesting subject, and more studies are warranted.

\section{The duality transformation} 
In the following we summarize the main steps in the duality transformation. For more details please refer to Section A of Appendix. We consider \Eq{Hs} defined on an infinite simple cubic lattice. Note that if one performs duality transformation on a finite lattice instead, twisted boundary conditions need to be imposed, under which the angle of the spins is allowed to change by $\Delta\theta$ across the boundary. Moreover, one needs to sum this angle from 0 to $2\pi$. This procedure is called ``orbifold''.
For analytic treatment, we replace the Boltzmann weight $$\exp\left[\beta J_1\cos(\theta_i-\theta_j)+\beta J_2 \cos2(\theta_i-\theta_j)\right]$$ by that of the Villain\cite{Villain_1975} model which has the same U(1) symmetry and captures the  $2\pi$ periodicity in $\theta_i$. The bond Boltzmann weight of the Villain model is given by, 
\be &&\exp[V(\theta_i-\theta_j)]=\sum_{m_{i,j}\in \mathbb{Z}} \Big\{\exp\left[-{K\over 2}(\theta_i-\theta_j-2\pi m_{i,j})^2\right]\nn&&+ \exp(-\Delta) \exp\left[-{K\over 2}(\theta_i-\theta_j-\pi-2\pi m_{i,j})^2\right]\Big\}.\nonumber \ee
In the above equation $K$ is the curvature of the potential around $\theta_i-\theta_j=2n\pi {\rm ~and~} (2n+1)\pi$.  
$\Delta$ reflects the energy difference between the local energy minimum $\theta_i-\theta_j=(2n+1)\pi$ and $\theta_i-\theta_j=2n\pi$ and the summation is over integers $m_{i,j}$ defined on links $\left< ij\right>$ .  For simplicity, we have assumed the curvature of the interaction energy around $\theta_i-\theta_j=2n\pi$ and  $\theta_i-\theta_j=(2n+1)\pi$ to be the same.

By Fourier transforming the Villain Boltzmann weight and integrating out $\theta_i$ variables,
the partition function reads
\be
Z
&&=\sum_{\{\nabla\cdot\v {\cal{L}}_{i}=0\}}\Tr_{\t}\prod_{\left< ij\right>}\exp\left(-\frac{{\cal{L}}_{ij}^2}{2K}-\Delta{\t_{ij}} -i\pi{\cal{L}}_{ij}\t_{ij}\right).\nn\label{eq:villain}
\ee
Here 
$\Tr_{\t}:=\sum_{\{\t_{ij}=0,1\}},$ ``$\nabla\cdot$'' denotes the lattice divergent 
and $\mathcal{L}_{ij}$ is an integer-valued field defined on link $\left<ij\right>$ with $\v {\cal{L}}_i=({\cal{L}}_{i,i+\hat{x}}, {\cal{L}}_{i,i+\hat{y}}, {\cal{L}}_{i,i+\hat{z}})$.

To solve the constraints $\nabla\cdot\v{\cal{L}}_i=0$, we introduce integer-valued variables $\cal{N}_{RR'}$ residing on the links of the dual lattice $\{R\}$, by $ {\cal{L}}_{ij}=\sum^{\circlearrowleft}_{\<RR'\>\in \square_{\<ij\>}}{\cal{N}}_{RR'}:={\cal{N}}_{R,R+\hat{x}}+{\cal{N}}_{R+\hat{x},R+\hat{x}+\hat{y}}-{\cal{N}}_{R,R+\hat{y}}-{\cal{N}}_{R+\hat{y},R+\hat{x}+\hat{y}}.$ Here $\square_{\<ij\>}$ is the dual plaquette whose center is the mid point of  $\<ij\>$ (see \Fig{dualrel}(a)). 
\begin{figure}[H]
  \centering
		\includegraphics[scale=0.25]{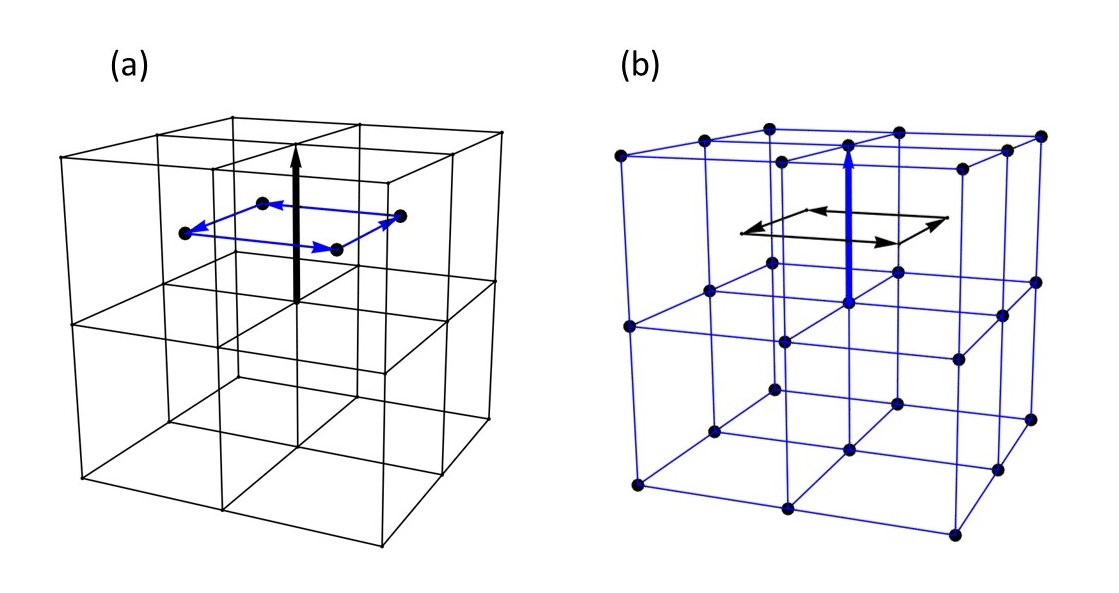}
		\caption{(a) The relation between ${\cal{L}}_{ij}$ (thick black arrow) and the ${\cal{N}}_{RR'}$ (blue arrows). The thin black lines are the links of the original lattice, and the large black dots are the dual lattice sites. (b) The relation between $\phi_{RR'}$ (thick blue arrow) and the $\t_{ij}$ (thin black arrows) in \Eq{cstt}. The large black dots and blue  lines are the sites and links on the dual lattice.}
		\label{dualrel}
	\end{figure}
$ {\cal{L}}_{ij}=\sum^{\circlearrowleft}_{\<RR'\>\in \square_{\<ij\>}}\cal{N}_{RR'}$ can be abbreviated as
$\v{\cal{L}}_i=\nabla\times \v{\cal{N}}_R.$
Under these definitions and by the virtue of Poisson's summation formula which changes the sum over $\v {\cal{N}}_R$ to the integral over a  continuous vector field $\v A_R$, the partition function reads 
\be
Z&&=\Tr_{\cal{M},\t}\int D[\v A_R]\prod_R\exp\Big[-\frac{(\nabla\times\v A_R)^2}{2K}\Big]\prod_{\<RR'\>}\nn&&\exp\Big[i(2\pi {\cal{M}}_{RR'}+\pi \phi_{RR'}) A_{RR'}\Big]\prod_{\<ij\>}\exp\left(-\Delta\t_{ij}\right),\label{gmmm}
\ee
where $\Tr_{{\cal M}}:=\prod_{\<RR'\>} \sum_{{\cal M}_{RR'}\in \mathbb{Z}}$. ${\cal{M}}_{RR'}$ and ${1\over 2}\phi_{RR'}$ are the vorticity of integer and half-integer vortices, respectively.  Here $\phi_{RR'}$ is a $\mathbb{Z}_2$-valued quantity defined as
\be
\phi_{RR'}=\sum_{\<ij\>\in \square_{\<RR'\>}}\t_{ij} \text{ mod }2.\label{cstt}
\ee

In Eq.(\ref{gmmm}) both integer vortices and half-integer vortices interact with a non-compact U(1) gauge field ($\v A$). However, there is an extra term $\prod_{\<ij\>}\exp\left(-\Delta\t_{ij}\right)$ in the partition function. It turns out this extra term gives rise  to a $\mathbb{Z}_2$ gauge field which only couples to the half-integer vortices.
To see that we first enforce \Eq{cstt} by a projection operator
\be\mathcal{P}:=\prod_{\<RR'\>}\frac{1}{2}\sum_{\eta_{RR'}=0,1}\left[
(-1)^{\phi_{RR'}}\times\prod_{\<ij\>\in \square_{\<RR'\>}}(-1)^{\t_{ij}}\right]^{\eta_{RR'}}\nn
\label{pjector}\ee
In Section A of the Appendix, between Eq. (A8) to Eq. (A17), we show that
$$
\mathcal{P}\prod_{\<ij\>} \exp\left(-\Delta\t_{ij}\right)\ra Z_{\mathbb{Z}_{\rm 2gauge}}\times\<W[\{\phi_{RR'}\}]\>
$$
where $Z_{\mathbb{Z}_2 \rm gauge}$ is the partition function of $\mathbb{Z}_2$ gauge theory and $\<W[\{\phi_{RR'}\}]\>$ is the expectation value of the Wilson loop\cite{Fradkin:1978aa,Kogut:1979aa}, namely,

\be
&&Z_{\mathbb{Z}_2\rm gauge}=\Tr_{\s} \exp\left( K_d\prod_{\left< RR'\right>\in \square_{\<ij\>}}\s_{RR'}\right)\nn
&&\<W[\{\phi_{RR'}\}]\>:=\Big\<\prod_{\<RR'\>}\s_{RR'}^{\phi_{RR'}}\Big\>_{\mathbb{Z}_{\rm 2gauge}}\label{w}
\ee
In \Eq{w} $\Tr_{\s}:=\prod_{\<RR'\>}\sum_{\s_{RR'}=0,1}$ and $\tanh K_d=\exp(-\Delta)$.
$\s_{RR'}=\pm 1$ are $\mathbb{Z}_2$ variables defined on dual links. Note in \Eq{w} only the $\s_{RR'}$ associated with dual links with $\phi_{RR'}\ne 0$ appears in the Wilson loop. The fact that those dual links form loops is because $\phi_{R,R+\hat{x}}+\phi_{R,R-\hat{x}}+\phi_{R,R+\hat{y}}+\phi_{R,R-\hat{y}}+\phi_{R,R+\hat{z}}+\phi_{R,R-\hat{z}}=0 ~(\text{mod }2).
$ More details for this part of duality transformation is given in Section A (from Eq. (A8) to Eq. (A17)) of the Appendix. 

The final partition function is given by
\be
&&Z=Z_{\mathbb{Z}_2{\rm gauge}}\Tr_{{\cal M}}\Tr_{\phi}\int D[\v A_R]\prod_R\exp\Big[-\frac{(\nabla\times\v A_R)^2}{2K}\Big]\nn&&\prod_{\<RR'\>}\exp\Big[i(2\pi {\cal{M}}_{RR'}+\pi\phi_{RR'}) A_{RR'}\Big]\<W[\{\phi_{RR'}\}]\>\label{gm12}
\ee
where
$\Tr_{\phi}:=\prod_{\<RR'\>} \sum_{\phi_{RR'}=0,1}.$ It can be seen that vortices in \Eq{gm12} interact with two gauge fields. Both integer and half-integer vortices interact with a non-compact U(1) gauge field \cite{Dasgupta:1981aa,Fisher:1989aa} $\v A_R$. In contrast, the half-integer vortices interact with a $\mathbb{Z}_2$ gauge field via the Wilson loop expectation value $\<W[\{\phi_{RR'}\}]\>$. In the confined phase of the $\mathbb{Z}_2$ gauge theory, the Wilson loop obeys the area law \cite{Kogut:1979aa,Fradkin:1979}. Under such condition the half-integer vortices are confined. Therefore at long wavelength only the integer vortices are present. Depending on the coupling strength to the U(1) gauge field, the integer vortex loops can be bound, or unbound. In the Former case there is long-range order in $\<\exp(i\theta_j)\exp(-i\theta_k)\>$ and   
$\<\exp(2i\theta_j)\exp(-2i\theta_k)\>$. In latter case both $\<\exp(i\theta_j)\exp(-i\theta_k)\>$ and   
$\<\exp(2i\theta_j)\exp(-2i\theta_k)\>$ decays exponentially.
If the Wilson loop obeys perimeter law  \cite{Kogut:1979aa,Fradkin:1979}, the half-integer vortices will be deconfined. Now both the integer and half-integer vortices interact only with the U(1) gauge field. If the interaction is strong enough to bind the half-integer vortex loops, the integer vortex loops will also be bound (because the integer vortices has twice the U(1) charge), in such case $\<\exp(i\theta_j)\exp(-i\theta_k)\>$ will exponential decay (due to the proliferation of the $\theta_i-\theta_j=\pi$ domain walls) while $\<\exp(2i\theta_j)\exp(-2i\theta_k)\>$ exhibit long-range order. If the  interaction with the U(1) gauge field is not strong enough to bind the fractional vortex loops, both
$\<\exp(i\theta_j)\exp(-i\theta_k)\>$ and $\<\exp(2i\theta_j)\exp(-2i\theta_k)\>$ will decay exponentially.

The above duality transformation can be generalized to situations where fractional vortices interact with a $\mathbb{Z}_n$ gauge field. In that case the generalized classical XY model read
\be
H_n=-J_1\sum_{\<ij\>}\cos(\theta_i-\theta_j)-J_n\sum_{\<ij\>}\cos n(\theta_i-\theta_j),\label{znXY}
\ee
for integer $n$. A similar duality transformation yields the following dual partition function
\be
&&Z_n=Z_{\mathbb{Z}_n{\rm gauge}}\Tr_{{\cal M}}\Tr_{\phi}\int D[\v A_R]\prod_R\exp\Big[-\frac{(\nabla\times\v A_R)^2}{2K}\Big]\nn&&\prod_{\<RR'\>}\exp\Big[i(2\pi {\cal{M}}_{RR'}+\frac{2\pi}{n}\phi_{RR'}) A_{RR'}\Big]\<W[\{\phi_{RR'}\}]\>,
\ee
where the vorticities of integer and fractional-vortices are ${\cal{M}}_{RR'}$ and $\frac{1}{n}\phi_{RR'}$, respectively. $\<W[\{\phi_{RR'}\}]\>$ is now evaluated in $\mathbb{Z}_n$ gauge theory. Similar to the $\mathbb{Z}_2$ case, there is a $\mathbb{Z}_n$ confinement-deconfinement transition.  For details of this analysis of the $\mathbb{Z}_n$ case please refer to Section B of the Appendix.

\section{Topological excitations}
\subsection{The Wilson and 't Hooft Loops of the $\mathbb{Z}_2$ Gauge Theory}
\begin{figure}[H]
	\centering
	\includegraphics[scale=0.14]{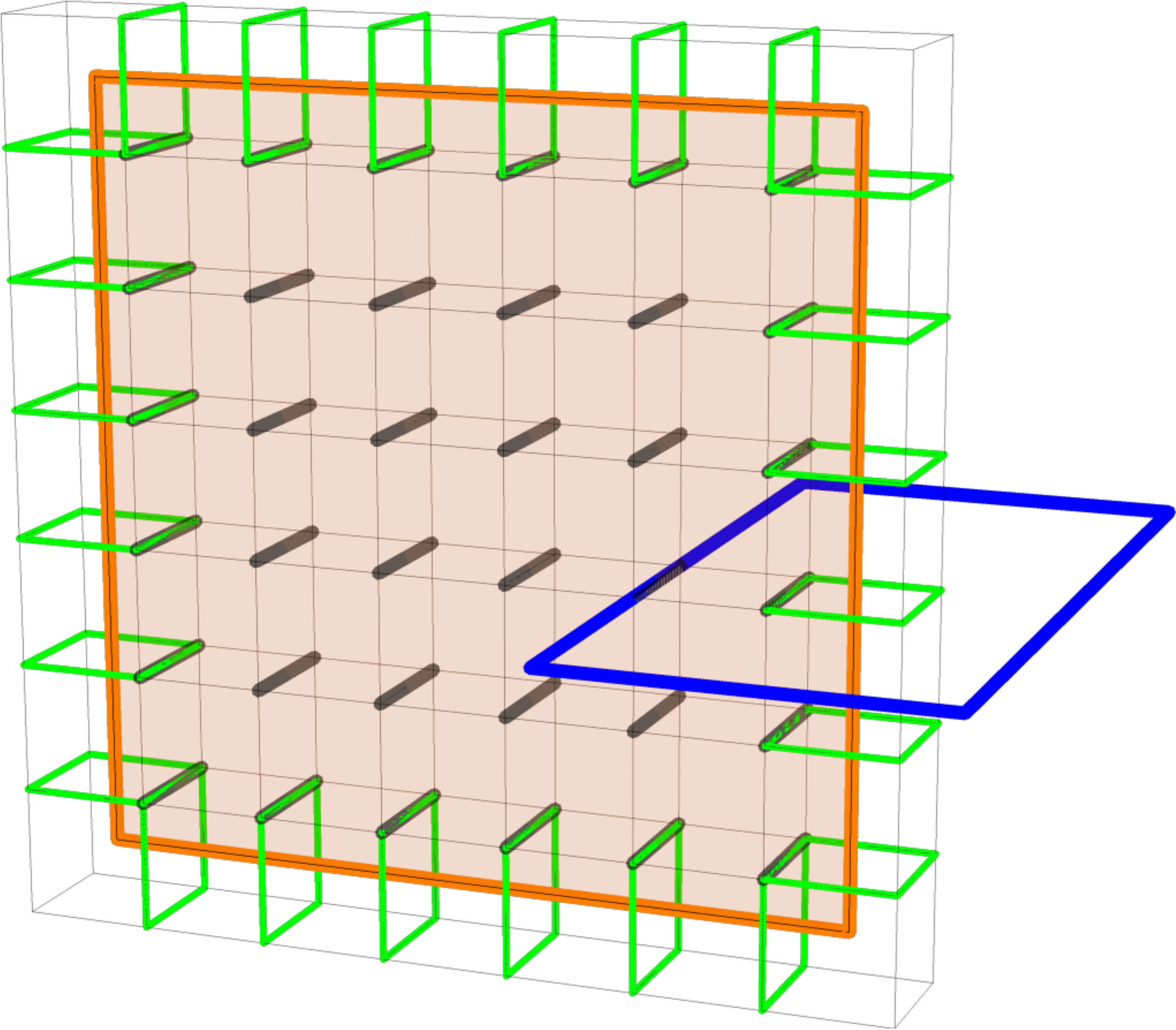}
	\caption{ Linking between Wilson loop and 't Hooft loop for $\mathbb{Z}_2$ gauge theory. Blue loop: the Wilson loop $\Gamma_W$. Orange loop: the 't Hooft loop $\Gamma_{T}$. Orange surface: a surface $\mathcal{A}_{\Gamma_{T}}$ that is enclosed by $\Gamma_{T}$ such that $\partial \mathcal{A}_{\Gamma_{T}} = \Gamma_{T}$. Green squares: the plaquettes pierced through by the 't Hooft loop. Thick black links: the bonds cut through by the chosen surface $\mathcal{A}_{\Gamma_{T}}$.}
	\label{loops}
\end{figure}

In the following we review  the Wilson and 't Hooft loops from the path integral point of view.
For a gauge theory, the Wilson loop is defined as the product of the gauge connections along a loop $\Gamma_W$ living on the dual  lattice. For $\mathbb{Z}_2$ gauge theory, this corresponds to the insertion of an operator\cite{Kogut:1979aa}
\begin{align}
	W_{\Gamma_W}(\{ \sigma_{RR'}\}) = \prod_{\langle R R'\rangle \in \Gamma_W} \sigma_{RR'}
\end{align}
\noindent This is drawn as the blue loop in \Fig{loops}.
The 't Hooft loop $\Gamma_T$, on the other hand, pierces through the plaquettes of the space-time lattice. An example is the orange loop in \Fig{loops}. 
The insertion of a 't Hooft loop  changes the signs of the flux of a plaquette if the plaquette is pierced through by $\Gamma_T$, namely
\begin{align}
&\prod_{\left< RR'\right>\in \square_{\<ij\>}}\s_{RR'}
	\rightarrow \Xi_{\G_T}\Big(\square_{\<ij\>}\Big)\prod_{\left< RR'\right>\in \square_{\<ij\>}}\s_{RR'}.
	\label{tHooft1}
\end{align}
Here $\Xi_{\G_T}\Big(\square_{\<ij\>}\Big)$ is $-1$ if $\square_{\<ij\>}$ is pierced through by $\Gamma_T$, and $+1$ otherwise. The plaquettes which are pierced by the 't Hooft loop are painted in green in \Fig{loops}.

When the 't Hooft loop is contractible, there is an alternative way to insert the 't Hooft loop. 
Due to the contractibility, one can find a surface $\mathcal{A}_{\Gamma_T}$ bounded by $\Gamma_T$, such that $\partial \mathcal{A}_{\Gamma_T} = \Gamma_T$. Such a surface will cut through some links in the space-time link lattice. The links cut through are drawn by thick black line in \Fig{loops}. It doesn't take long to convince oneself that the insertion of 't Hooft loop corresponds to flipping the signs of the link variables that are cut through by $\mathcal{A}_{\Gamma_T}$, namely
\be
	\sigma_{RR'}\rightarrow\Upsilon_{\mathcal{A}_{\Gamma_T}}\Big(\<RR'\>\Big)~ \sigma_{RR'}.\label{tHooft2}\ee
where $\Upsilon_{\mathcal{A}_{\Gamma_T}}\Big(\<RR'\>\Big)=-1$ if $\<RR'\>$ is intersected by $\mathcal{A}_{\Gamma_T}$, otherwise it is $+1$.
\noindent One can easily show that the result only depends on $\Gamma_T$ but not on the specific choice of $\mathcal{A}_{\Gamma_T}$.

One reason why Wilson and 't Hooft loops are interesting is that they correspond to the world lines of the quasiparticles in the topological ordered phase \cite{wen1990,kitaev2003} (the deconfined phase) realized when $K_d$ in \Eq{w} is sufficiently large. 
As we shall see in the following,  these two types of quasi-particle has mutual $-1$ statistics.

\subsection{Mutual Statistics of Wilson and 't Hooft Loops}
Consider a $\mathbb{Z}_2$ gauge theory with the insertion of a Wilson loop. We would like to know how the partition function would change when a 't Hooft loop which has non-trivial linking with the Wilson loop is inserted. One such example is drawn in \Fig{loops}.
The partition function with a single Wilson loop $\Gamma_W$ insertion, namely, $Z_{\mathbb{Z}_2\rm gauge}\left[ \Gamma_W \right]$, is given by
\be
&&Z_{\mathbb{Z}_2\rm gauge}\left[ \Gamma_W \right]=\nn &&\Tr_\s \left( \prod_{\langle R R'\rangle \in \Gamma_W} \sigma_{RR'} \right) \prod_{\square_{\<ij\>}}\exp\left[ K_d\prod_{\left< RR'\right>\in \square_{\<ij\>}}\s_{RR'}\right].\nn\label{singlew}
\ee
On the other hand, in the presence of an additional 't Hooft loop we replace \Eq{singlew} by
\be
&&Z_{\mathbb{Z}_2\rm gauge}\left[ \Gamma_W, \Gamma_T \right]=\Tr_\s\left( \prod_{\langle R R'\rangle \in \Gamma_W} \sigma_{RR'} \right)\times\nn&& \prod_{\square_{\<ij\>}}\exp\left[ K_d\prod_{\left< RR'\right>\in \square_{\<ij\>}}\Upsilon_{\mathcal{A}_{\Gamma_T}}\Big(\<RR'\>\Big)~ \sigma_{RR'}\right].\label{two_loops}
\ee

For a specific $\mathcal{A}_{\Gamma_T}$, 
\Eq{tHooft2} is a one-to-one mapping of the $\mathbb{Z}_2$ link variables. Because $\left\{ \sigma_{RR'} \right\}$ are dummy variables to be summed over in \Eq{two_loops}, we can do the replacement of \Eq{tHooft2} in \Eq{two_loops}, which leads to 
\begin{align}
	Z_{\mathbb{Z}_2\rm gauge}\left[ \Gamma_W, \Gamma_T \right] = (-1)^{n_{\text{linking}}} Z_{\mathbb{Z}_2\rm gauge}\left[ \Gamma_W \right],
\end{align} Here $n_{\text{intercept}}$ is the number of times $\Gamma_W$ intercepts with $\mathcal{A}_{\Gamma_T}$, i.e., the linking number between $\Gamma_W$ and $\Gamma_T$. Physically this means that if the additional 't Hooft loop has non-trivial linking number \cite{witten1989} (mod 2) with the Wilson loop, the partition function would gain an additional $-1$ sign. In the literature the Wilson/'t Hooft loop corresponds to the world line of the ``$e$/$m$ particle'' of $\mathbb{Z}_2$ topological order \cite{kitaev2003}. The result above implies that $e$ and $m$ particles have mutual $-1$ statistics\cite{kitaev2003} with respect to each others.

\section{Quantum formulation}
\begin{figure}[H]
  \centering
		\includegraphics[scale=0.35]{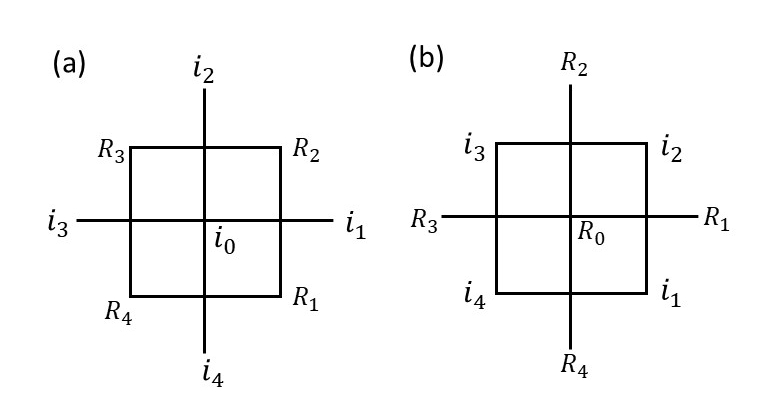}
		\caption{(a) The relation between the links on the original and dual lattice in \Eq{qdual}. (b) The relation between the plaquettes in the original lattice and the stars in the dual lattice.}
		\label{qmdual}
	\end{figure}
To see the topological aspects of this model, we map the 3D classical model into a $(2+1)$D quantum model defined on 2D square lattice. The quantum Hamiltonian reads
\be
\hat{H}&=&\sum_{\left<ij\right>}-J_1\cos(\theta_i-\theta_j)-J_2\cos(2\theta_i-2\theta_j)\nonumber\\
&+&\sum_i\frac{1}{2(J_1+4J_2)}n_i^2-\frac{\ln\coth J_1}{2}\cos\pi n_i,\label{Quantum}
\ee
where $[\theta_j,n_k]=i\delta_{jk}$.
For details of the derivation please refer to Section C of the Appendix. In $J_2=0$ limit (corresponding to $\Delta\rightarrow +\infty$), there is a paramagnetic to ferromagnetic phase transition when $J_1$ is tuned from $0$ to $+\infty$, with U(1) symmetry spontaneously breaking and order parameter $e^{i\theta_i}$. In $J_1=0$ limit (corresponding to $\Delta=0$), there is another paramagnetic to ferromagnetic phase transition when $J_2$ is tuned from $0$ to $+\infty$, with U(1) symmetry spontaneously breaking and order parameter $e^{i2\theta_i}$. What's more interesting is the confinement-deconfinement transition as a function of $\Delta$ (or $K_d$), while the interaction with the U(1) gauge field is infinitely strong ($K\ra\infty$ in \Eq{gm12}). This limit corresponds $J_2\ra\infty$ limit in \Eq{Hs} and \Eq{Quantum}, where the quantum model has following Hamiltonian and Hilbert space
\be
&&\lim_{J_2\rightarrow +\infty}\hat{H}= -J_1\sum_{\left<ij\right>}\cos(\theta_i-\theta_j)-\frac{\ln\coth J_1}{2}\sum_i\cos\pi n_i.\nn
&&\text{Hilbert space constraint:~} \theta_i-\theta_j=k_{ij}\pi, k_{ij}\in\mathbb{Z}. 
\label{j2div}\ee 
Define a $\mathbb{Z}_2$ variable on the dual links of the square lattice: \be\s^x_{R_1R_2}=e^{i(\theta_{i_0}-\theta_{i_1})}\Rightarrow \cos(\theta_{i_0}-\theta_{i_1})=\s^x_{R_1R_2}\label{qdual}\ee (see \Fig{qmdual}(a)).  
Because $$e^{i(\theta_{i_1}-\theta_{i_2})}e^{i(\theta_{i_2}-\theta_{i_3})}e^{i(\theta_{i_3}-\theta_{i_4})}e^{i(\theta_{i_4}-\theta_{i_1})}=1$$ it follows that \be\s^x_{R_0R_1}\s^x_{R_0R_2}\s^x_{R_0R_3}\s^x_{R_0R_4}:=\prod_{R_i\in\text{star of~} R_0}\s^x_{R_0R_i}=1\nn\label{gtc}\ee (see \Fig{qmdual}(b)).
Since $\cos\pi n_{i_0}\ket{\theta_{i_0}}=\ket{\theta_{i_0}+\pi}$, it flips the sign of $\s^x_{R_1R_2},\s^x_{R_2R_3},\s^x_{R_3R_4}, \s^x_{R_4R_1}$ in \Fig{qmdual}(a). Therefore $\cos \pi n_i\ra \prod_{\left<RR\right>\in \square_i}\s_{RR'}^z.$ Thus Hamiltonian reads
\be
&&\lim_{J_2\rightarrow +\infty}\hat{H}=-J_1\sum_{\left<RR^\prime\right>}\s^x_{RR^\prime}-\frac{\ln\coth J_1}{2}\sum_{\square}\prod_{\left<RR^\prime\right>\in\square}\s_{RR'}^z.\nn
\label{Igh}\ee
\Eq{Igh} and \Eq{gtc} are the quantum Hamiltonian, and the Hilbert space constraint of the $\mathbb{Z}_2$ gauge theory\cite{Fradkin:1978aa}. The confinement-deconfinement transition occurs as one tunes $J_1$ from infinity to zero. Under the constraint in \Eq{j2div},  the Hilbert space is the direct product of that spanned by $|\theta\>$ on a reference single site, and the Hilbert space of the $\mathbb{Z}_2$ gauge theory. The $\theta$ degrees of freedom reproduces the continuous degeneracy in the $\<e^{i2\theta_i}\>\ne 0$ phase due to the U(1) symmetry of \Eq{j2div}. Thus, the deconfined phase possesses simultaneously $\<e^{i2\theta_i}\>\ne 0$ and topological degeneracy on a torus when using the orbifold boundary conditions.

\section{Conclusion}

In conclusion, we generalize the 2D classical model proposed in Ref. \cite{Lee:1985aa} to 3D and show the emergence of the $\mathbb{Z}_2$ gauge theory from a model without any built-in gauge structures. The deconfined phase of this $\mathbb{Z}_2$ gauge theory shows a coexistence of topological order and conventional long-range order. By a further generalization of our model, we also construct a simple model realizing $\mathbb{Z}_n$ gauge theory. The model \Eq{Hs} can be realized in cold atom experiments in either 2D or 3D. It remains an interesting question to detect the topological signatures of our model in experiments.\\

{\bf Acknowledgments:} ZQG acknowledges Congjun Wu for a discussion. This work was primarily funded by the U.S. Department of Energy, Office of Science, Office of Basic Energy Sciences, Materials Sciences and Engineering Division under Contract No. DE-AC02-05-CH11231 (Theory of Materials program KC2301). This research is also funded in part by the Gordon and Betty Moore Foundation.\\

\bibliographystyle{apsrev4-1}
\bibliography{bibs}

\onecolumngrid
\appendix

\section{The duality transformation that leads to partition function Eq.(7) of the main text}
We start from Eq.(2) of the main text
\be
Z=\sum_{\{\nabla\cdot\v {\cal{L}}_{i}=0\}}\Tr_{\t}\prod_{\left< ij\right>}\exp\left(-\frac{{\cal{L}}_{ij}^2}{2K}-\Delta{\t_{ij}} -i\pi{\cal{L}}_{ij}\t_{ij}\right),\label{part}
\ee
where ${\cal{L}}_{ij}$ are integer-valued link variables, and $\v {\cal{L}}_i:=({\cal{L}}_{i,i+\hat{x}}, {\cal{L}}_{i,i+\hat{y}}, {\cal{L}}_{i,i+\hat{z}})$. The lattice divergent-less constraint $\nabla\cdot\v {\cal{L}}_{i}=0$ is solved by
\be\v{\cal{L}}_i=\nabla\times \v{\cal{N}}_R\label{curl}.\ee Here $\v{\cal{N}}_R$ is an integer-valued vector field on the dual lattice $\{R\}$. Plug \Eq{curl} into \Eq{part} we obtain
\be
Z=\Tr_{\tau}\sum_{\{\v{{\cal{N}}}_R\}}\prod_R\exp\Big[-\frac{(\nabla\times\v{\cal{N}}_R)^2}{2K}\Big]\prod_{ij}\exp\Big[-\Delta\t_{ij}+i\pi (\nabla\times\v{\cal{N}}_R)_{ij}\t_{ij}\Big]. \notag
\ee
After applying the Poisson's summation formula, we change the sum over $\v {\cal{N}}_R$ to the integral over a  continuous vector field $\v A_R$, which plays the role of a non-compact U(1) gauge field,  and the partition function becomes 
\be
Z=\Tr_{\cal{M},\t}\int D[\v A_R]\prod_R\exp\Big[-\frac{(\nabla\times\v A_R)^2}{2K}\Big]\prod_{\<RR'\>}\exp\Big[i(2\pi {\cal{M}}_{RR'}+\pi \Phi_{RR'}) A_{RR'}\Big]\prod_{\<ij\>}\exp\left(-\Delta\t_{ij}\right),\label{gmm}
\ee
where $\Tr_{{\cal M}}:=\prod_{\<RR'\>} \sum_{{\cal M}_{RR'}\in \mathbb{Z}}.$ In \Eq{gmm} \be \Phi_{RR'}=\sum^{\circlearrowleft}_{\<ij\>\in \square_{\<RR'\>}}\t_{ij}=\t_{i,i+\hat{x}}+\t_{i+\hat{x},i+\hat{x}+\hat{y}}-\t_{i,i+\hat{y}}-\t_{i+\hat{y},i+\hat{x}+\hat{y}}\label{orien}\ee takes integer value from $-4$ to $4$. The oriented sum is taken over the four links around the plaquette $\square_{\<RR'\>}$ in the original lattice which is perpendicular to $\<RR'\>$ (see \Fig{dualrel}(b)).

\begin{figure}[H]
  \centering
		\includegraphics[scale=0.25]{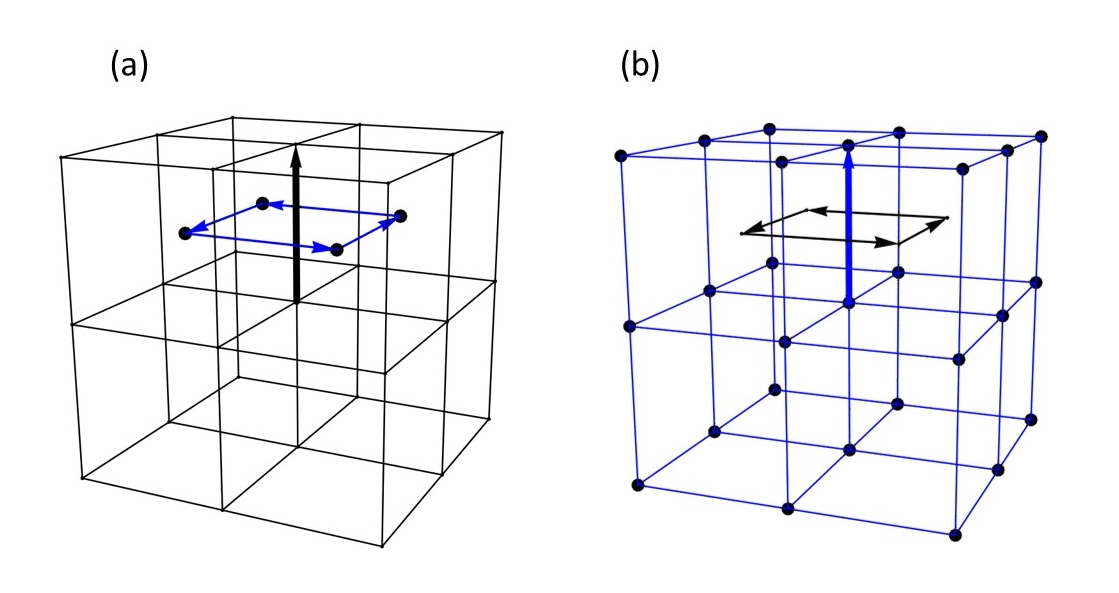}
		\caption{(a) The relation between ${\cal{L}}_{ij}$ (thick black arrow) and the ${\cal{N}}_{RR'}$ (blue arrows) in \Eq{curl}. The thin black lines are the links of the original lattice, and the large black dots are the dual lattice sites. (b) The relation between $\Phi_{RR'}$ (thick blue arrow) and the $\t_{ij}$ (thin black arrows) in \Eq{orien}. The large black dots and blue  lines are the sites and links on the dual lattice.}
		\label{dualrel}
	\end{figure}
	
However, due to $\Tr_{\cal{M}}$ in \Eq{gmm}, where ${\cal{M}}_{RR'}$ is integer valued on each dual link,  the part of $\Phi_{RR'}$ which is a multiple of $2$ can be absorbed into ${\cal{M}}_{RR'}$. The remaining  part of $\Phi_{RR'}$,
\be&&\phi_{RR'}=\sum_{\<ij\>\in \square_{\<RR'\>}}\t_{ij} \text{ mod }2,\label{fxx}\ee
 is a $\mathbb{Z}_2$-valued quantity. As a result, \Eq{fxx} is equivalent to \be
&&(-1)^{\phi_{RR'}}=\prod_{\<ij\>\in \square_{\<RR'\>}}(-1)^{\t_{ij}}.\label{cst}\ee
In terms of $\phi_{RR'}$ the partition function is given by
\be
Z=\Tr_{\cal{M},\t}\int D[\v A_R]\prod_R\exp\Big[-\frac{(\nabla\times\v A_R)^2}{2K}\Big]\prod_{\<RR'\>}\exp\Big[i(2\pi {\cal{M}}_{RR'}+\pi \phi_{RR'}) A_{RR'}\Big]\prod_{\<ij\>}\exp\left(-\Delta\t_{ij}\right),
\ee

As stated in the main text, the constraints \Eq{cst} can be imposed through a projection operator defined as
\be
\mathcal{P}:=\prod_{\<RR'\>}\frac{1}{2}\sum_{\eta_{RR'}=0,1}\left[
(-1)^{\phi_{RR'}}\times\prod_{\<ij\>\in \square_{\<RR'\>}}(-1)^{\t_{ij}}\right]^{\eta_{RR'}}.
\ee
After inserting $\mathcal{P}$  into the partition function,
$\t_{ij}$ and $\phi_{RR'}$ become unconstrained  variables which can be summed over independently. If we define
\be
\mathcal{B}[\{\phi_{RR'}\}]:= \Tr_{\eta}\Tr_{\t}\left(\mathcal{P}\prod_{\left< ij\right>}\exp\left(-\Delta\t_{ij}\right)\right)\label{a3d}
\ee
where $\Tr_{\eta}:=\prod_{\<RR'\>}\sum_{\eta_{RR'}=0,1}.$ It is straightforward to show that after performing $\Tr_{\t}$ in \Eq{a3d}, 
\be
\mathcal{B}[\{\phi_{RR'}\}]\propto \Tr_{\eta}\left(\prod_{\<RR'\>} (-1)^{\eta_{RR'}\phi_{RR'}}\right)\prod_{\square_{\<ij\>}}F_2\left[\{\eta_{RR'}\}\right],\label{g3}
\ee
where
\be
F_2\left[\{\eta_{RR'}\}\right]=\exp\left[K_d\cos\left(\pi\sum_{\left< RR'\right>\in \square_{\<ij\>}}\eta_{RR'}\right)\right].\label{lgK}\ee
In \Eq{lgK}  \be\tanh K_d=\exp(-\Delta)\label{gcoup}\ee is the coupling constant in the dual lattice gauge theory.
By introducing a $\mathbb{Z}_2$ variable on each link  $$\s_{RR'}:=(-1)^{\eta_{RR'}},$$ $$F_2\left[\{\eta_{RR'}\}\right]\ra  \tilde{F}_2\left[\{\s_{RR'}\}\right]$$ where 
\be
\tilde{F}_2\left[\{\s_{RR'}\}\right]=\exp\left( K_d\prod_{\left< RR'\right>\in \square_{\<ij\>}}\s_{RR'}\right)\label{gauge}
\ee
is the Boltamann weight of $\mathbb{Z}_2$ gauge theory. Therefore 
\be
\mathcal{B}[\{\phi_{RR'}\}]=\Tr_{\s}\left\{\s_{R_1R_1'}...\s_{R_NR_N'}\exp\left( K_d\prod_{\left< RR'\right>\in \square_{\<ij\>}}\s_{RR'}\right)\right\},\label{eq:gauge}
\ee
where $\<R_1R'_1\>...\<R_NR'_N\>$ label are the dual links for which $\phi_{RR'}\ne 0$. The $\mathcal{B}[\{\phi_{RR'}\}]$ given by 
\Eq{eq:gauge} is proportional to
the expectation value of the Wilson loop\cite{Fradkin:1978aa,Kogut:1979aa} in the $\mathbb{Z}_2$ gauge theory, namely,
\be
\mathcal{B}[\{\phi_{RR'}\}]=Z_{\mathbb{Z}_2{\rm gauge}}\<W[\{\phi_{RR'}\}]\>\label{Wilson}\ee where 
\be &&Z_{\mathbb{Z}_2{\rm gauge}}=\Tr_{\s}\exp\left(K_d\prod_{\< RR'\>\in \square_{\<ij\>}}\s_{RR'}\right)
\ee
and
\be \<W[\{\phi_{RR'}\}]\>:=\Big\<\prod_{\<RR'\>}\s_{RR'}^{\phi_{RR'}}\Big\>_{\mathbb{Z}_2{\rm gauge}}
\ee
\section{The duality transformation that lead to the U(1) and $\mathbb{Z}_n$ gauge theory}

We consider the following classical Hamiltonian
\be
H=\sum_{\left< ij\right>}\left[-J_1 \cos(\theta_i-\theta_j) - J_n \cos n(\theta_i - \theta_j)\right],\label{eq:xy}
\ee on an infinite  simple cubic lattice, 
where $J_1$ and $J_n$ are positive coupling constant parameters, and $n>2$ is an integer. 
The partition function reads
\be
Z=\int \prod_i d\theta_i\prod_{\left< ij\right>}\exp( \beta J_1 \cos(\theta_i-\theta_j)+\beta J_n \cos n(\theta_i - \theta_j)),
\ee
The Villain form\cite{Villain_1975} of the Boltzmann weight for each link $\<ij\>$ is given by
\be
\exp{V(\theta_i-\theta_j)}=\sum_{m_{i,j}=-\infty }^{\infty}\sum_{\t_{ij}=0}^{n-1}\exp\Big[-\Delta(\t_{ij})-\frac{K}{2}\Big(\theta_i-\theta_j-2\pi m_{i,j}-\frac{2\pi}{n}\t_{ij}\Big)^2\Big],
\label{vl}
\ee
and $$\Delta(\t_{ij})=K\left(1-\cos \frac{2\pi \t_{ij}}{n}\right)$$ as a function of $\tau_{ij}$ measuring the energy difference between local minima at $\theta_i-\theta_j=2k\pi+\frac{2\pi \t}{n}$ and $\theta_i-\theta_j=2k\pi$. The Fourier transform of the Boltzmann weight in \Eq{vl} is given by (for simplicity we temporarily neglect $i,j$ indices)
\be
C_\mathcal{L}&=&\int_0^{2\pi}d\Delta\theta \exp{V[\Delta\theta]}
\exp(-i\mathcal{L}\Delta\theta)
=\text{Const.}\times \exp\left(-\frac{\mathcal{L}^2}{2K}\right)\sum_{\t=0}^{n-1}\exp\left(-\Delta(\t) -i\frac{2\pi}{n}\mathcal{L}\t\right).
\ee
Here $\mathcal{L}$ is an integer, defined for each link, and $\Delta\theta$ denotes the angle difference between site $i$ and $j$. Therefore, the partition function becomes
\be
Z&&=\int \prod_id\theta_i\sum_{\{\mathcal{L}_{ij}\}}\prod_{\left< ij\right>}\exp\left(-\frac{\mathcal{L}_{ij}^2}{2K}+i\mathcal{L}_{ij}(\theta_i-\theta_j)\right)\sum_{\t_{ij}=0}^{n-1}\exp\left(-\Delta({\t_{ij})} -i\frac{2\pi}{n}\mathcal{L}_{ij}\t_{ij}\right)\nn
&&=\sum_{\{\nabla\cdot \v {\mathcal{L}}_{i}=0\}}\Tr_{\t}\prod_{\left< ij\right>}\exp\left(-\frac{\mathcal{L}_{ij}^2}{2K}-\Delta({\t_{ij})} -i\frac{2\pi}{n}\mathcal{L}_{ij}\t_{ij}\right),\label{eq:villain}
\ee
where  $\Tr_{\t}:=\sum_{\{\t_{ij}=0,...,n-1\}}.$
In passing to the second line  of \Eq{eq:villain} we have integrated over $\theta_i$ which results in  the constraint \be \nabla\cdot \v {\mathcal{L}}_i=0.\label{divf}\ee Here $\nabla\cdot$ is the lattice divergence operator, and $\v {\mathcal{L}}_i$ is the vector formed by the near-neighbor $\mathcal{L}_{ij}$, i.e., 
$\v {\mathcal{L}}_i=(\mathcal{L}_{i,i+\hat{x}}, \mathcal{L}_{i,i+\hat{y}}, \mathcal{L}_{i,i+\hat{z}})$.\\


Like in Section I, we solve the constraints $\nabla\cdot\v {\mathcal{L}}_i=0$ by letting $\v {\mathcal{L}}_i=\nabla\times\v {\mathcal{N}}_{R}$, and use the Poisson's summation formula to convert the sum over the integer-valued vector field $\v {\mathcal{N}}_{R}$ to the integration over a non-compact gauge field ${\v A}_R$ \cite{Dasgupta:1981aa,Fisher:1989aa}  
\be
Z=\Tr_{\mathcal{M},\t}\int D[{\v A}_R]\prod_R\exp\Big[-\frac{(\nabla\times{{\v A}}_R)^2}{2K}\Big]\prod_{\<RR^\prime\>}\exp\Big[i(2\pi \mathcal{M}_{RR^\prime}+\frac{2\pi}{n}\phi_{RR^\prime}) {{A}}_{RR^\prime}\Big]\exp\left(-\Delta(\t_{ij})\right).\label{gm}
\ee
Here \be\phi_{RR^\prime}=(\t_{i,i+\hat{x}}+\t_{i+\hat{x},i+\hat{x}+\hat{y}}-\t_{i,i+\hat{y}}-\t_{i+\hat{y},i+\hat{x}+\hat{y}})~\text{mod}~n.\label{phi}\ee 
\\


\subsection{The emergence of the $\mathbb{Z}_n$ gauge field}
To enforce \Eq{phi}, we introduce the projection operator
\be
\mathcal{P}:=\prod_{\<RR^\prime\>}\frac{1}{n}\sum_{\eta_{RR^\prime}=1}^{n-1}\left(\omega^{\phi_{RR^\prime}}\prod^{\circlearrowleft}_{\<ij\>\in\square_{\< RR^\prime\>}}\omega^{-\t_{ij}}\right)^{\eta_{RR^\prime}},
\ee
where $$\omega=e^{i 2\pi/n},$$
and $\square_{\< RR^\prime\>}$ is the plaquette on the original lattice intersecting the dual link $\<RR^\prime\>$. After inserting this projector operator, 
$\t_{ij}$ and $\phi_{RR'}$ become independent variables ready to be traced over. If we define
\be
&&\mathcal{B}[\{\phi_{RR^\prime}\}] 
:=\Tr_{\eta}\left(\prod_{\<RR^\prime\>}\omega^{\eta_{RR^\prime}\phi_{RR^\prime}}\right)\Tr_{\t}\prod_{\left< ij\right>}\exp\left(-\Delta(\t_{ij})\right)\prod^{\circlearrowleft}_{\<ij\>\in \square_{\<RR'\>}}\omega^{-\t_{ij}\eta_{RR^\prime}}\label{a3dd}
\ee
it is straightforward to show that 
after the $\Tr_{\t}$, up to a constant, 
\be
\mathcal{B}[\{\phi_{RR^\prime}\}]\propto\Tr_{\eta}\left(\prod_{\< RR^\prime\>} \omega^{\eta_{RR^\prime}\phi_{RR^\prime}}\right)\prod_{\square_{\<ij\>}}F_n\left[\left\{\sum^{\circlearrowleft}_{\left< RR^\prime\right>\in \square_{\<ij\>}}\eta_{RR^\prime}\right\}\right]\label{gg3}\ee
where $\square_{\<ij\>}$ denotes the plaquette on the dual lattice which intersects $\<ij\>$, and
\be
&&F_n\left[\chi_{ij}\right]:=e^{-K}\sum_{k_{ij}=-\infty}^{+\infty}I_{\nu(\chi_{ij};k_{ij})}(K)~~{\rm where}~~\nu(\chi_{ij};k_{ij}):=\chi_{ij}-nk_{ij},\label{ggg33}\ee
and
\be \chi_{ij}= \sum^{\circlearrowleft}_{\left< RR^\prime\right>\in \square_{\<ij\>}}\eta_{RR^\prime}=\eta_{R,R+\hat{x}}+\eta_{R+\hat{x},R+\hat{x}+\hat{y}}-\eta_{R,R+\hat{y}}-\eta_{R+\hat{y},R+\hat{x}+\hat{y}}.\label{gg33}
\ee
In \Eq{ggg33}, $I_\nu(x)$ is the modified Bessel function of the first kind \cite{mathbook}.  According to the asymptotic expansion of the Bessel function, for large $K$, $$I_\nu (K)\sim \frac{1}{\sqrt{2\pi K}} \exp \left(\sqrt{K^2+\nu^2}-\frac{\nu^2}{K}\right)\sim \frac{1}{\sqrt{2\pi K}} \exp \left(K-\frac{\nu^2}{2K}\right)$$ hence
\be
F_n\left[\chi_{ij}\right]
\sim\sum_{k_{ij}=-\infty}^{+\infty}\exp\left[-\frac{(\chi_{ij}-nk_{ij})^2}{2K}\right]\sim\exp\left[K_d\cos\left(\frac{2\pi}{n}\chi_{ij}\right)\right],
\label{lagK}
\text{ where } K_d=\frac{n^2}{4\pi^2 K}.\label{kd}\ee
By introducing a $\mathbb{Z}_n$ variables on the links of the dual lattice $$q_{RR^\prime}:=\omega^{\eta_{RR^\prime}}$$ 
\be
F_n\left[\chi_{ij}\right]\ra \tilde{F}_n\left[\tilde{\chi}_{ij}\right]=\exp \left[K_d~\Re(\tilde{\chi}_{ij})\right]~~{\rm where~} \tilde{\chi}_{ij}=\prod^{\circlearrowleft}_{\left< RR^\prime\right>\in \square_{\<ij\>}}q_{RR^\prime}.
\label{gaugef}\ee
Note that $\tilde{\chi}_{ij}$ in \Eq{gaugef} is the plaquette flux in the  $\mathbb{Z}_n$ gauge theory and $\tilde{F}_n\left[\tilde{\chi}_{ij}\right]$ is its Boltzmann weight. Therefore
\be
\mathcal{B}[\{\phi_{RR^\prime}\}]=\Tr_q\Big\{q_{R_1R_1^\prime}^{\phi_{R_1R_1^\prime}}...q_{R_NR_N^\prime}^{\phi_{R_NR_N^\prime}}\prod_{\square_{\<ij\>}}\tilde{F}_n\left[\tilde{\chi}_{ij}\right]\Big\}.\label{eq:gauge1}
\ee
In \Eq{eq:gauge1} $\< R_1R_1^\prime\> ,...,\< R_{N}R_N^\prime\>$ are the dual links where $\phi_{R_iR_i^\prime}\ne 0$. The $\mathcal{B}[\{\phi_{RR^\prime}\}]$ given by
\Eq{eq:gauge1} is proportional to the expectation value of the Wilson loop in the $\mathbb{Z}_n$ gauge theory\cite{Kogut:1979aa}, namely,
\be
\mathcal{B}[\{\phi_{RR^\prime}\}]=Z_{\mathbb{Z}_n{\rm gauge}}\times W[\{\phi_{RR^\prime}\}]\label{Wilson}\ee where
\be Z_{\mathbb{Z}_n{\rm gauge}}=\Tr_q\prod_{\square_{\<ij\>}}\tilde{F}_n\left[\tilde{\chi}_{ij}\right],\quad 
W[\{\phi_{RR^\prime}\}:=\Big\<\prod_{\<RR'\>}q_{RR^\prime}^{\phi_{RR^\prime}}\Big\>_{\mathbb{Z}_n{\rm gauge}}.\label{wilson2}\ee
 Here $q_{R_1R_1^\prime} ,...,q_{R_{N}R_N^\prime}$ satisfy the constraint 
\be
&&q_{R,R+\hat{x}}\cdot q_{R,R-\hat{x}}\cdot q_{R,R+\hat{y}}\cdot q_{R,R-\hat{y}}\cdot q_{R,R+\hat{z}}\cdot q_{R,R-\hat{z}}=1\nn&& {\rm and~} q_{RR^\prime}=q_{RR^\prime}^*,
\ee
which enforces the continuity of the fractional vortex loops. The final partition function is given by
\be
Z=Z_{\mathbb{Z}_n{\rm gauge}}\times\Tr_{\phi,\mathcal{M}}\int D[\v {{A}}_R]\prod_R\exp\Big[-\frac{(\nabla\times\v {{A}}_R)^2}{2K}\Big]\prod_{\<RR^\prime\>}\exp\Big[i(2\pi \mathcal{M}_{RR^\prime}+\frac{2\pi}{n}\phi_{RR^\prime}) {{A}}_{RR^\prime}\Big]W[\{\phi_{RR^\prime}\}].\label{gm1}
\ee

In the special case of $n=2$ we recover the results of Section I.\\

\section{The Quantum Hamiltonian for $n=2$}

In this section, we show  the 2D quantum Hamiltonian,
\be
\hat{H}=\sum_{\left< ij\right>}-J_1\cos(\theta_i-\theta_j)-J_2\cos 2(\theta_i-\theta_j)+\sum_i\frac{U}{2}n_i^2-\Gamma\cos\pi n_i,
\ee
with $[\theta_j,n_k]=i\delta_{jk}$ is equivalent to the 3D classical XY Eq.(1) of the main text. \\ 
\be
Z&=&\Tr \left[e^{-\beta \hat{H}}\right] 
=\sum_{\{\theta_i^\t\}}\prod_\t\bra{\{\theta_i^\t\}}e^{-\epsilon \hat{H}}\ket{\{\theta_i^{\t+1}\}}\nn
&=&\sum_{\{\theta_i^\t\}}\exp\left(\sum_{\left< ij\right>}\sum_{\t}\epsilon J_1\cos(\theta_i^\t-\theta_j^\t)+\epsilon J_2\cos 2(\theta_i^\t-\theta_j^\t)\right)\prod_{i,\t}\bra{\theta_i^\t}\exp\left(-\frac{\epsilon U}{2}n_i^2 +\epsilon\Gamma\cos\pi n_i\right)\ket{\theta_i^{\t+1}},
\ee
where $\epsilon=\frac{\beta}{N}$ and $\prod_\t$ is the product over imaginary time slices $\{\tau\}$. Insert $\mathbb{I}=\sum_n\ket{n}\bra{n}$ 
we obtain (up to a constant)
\be
\bra{\theta_i^\t}\exp\left(-\frac{\epsilon U}{2}n_i^2 +\epsilon\Gamma\cos\pi n_i\right)\ket{\theta_i^{\t+1}}
=\exp\left(\frac{\Delta}{2}\cos\left(\theta_i^\t-\theta_i^{\t+1}\right)+\frac{1}{4}\left(\frac{1}{\epsilon U}-\frac{\Delta}{2}\right)\cos2\left(\theta_i^\t-\theta_i^{\t+1}\right)\right),
\ee
where $\tanh\frac{\Delta}{2}=e^{-2\epsilon\Gamma}$, and we have used the fact that $e^{\epsilon\Gamma\cos \pi n}\propto 1+e^{-\Delta-i\pi n}$ for integer $n$. Therefore, the quantum partition function is given by
\be
Z&=&\sum_{\{\theta_i^\t\}}\prod_\t\exp\sum_{\left< ij\right>}\left(\epsilon J_1\cos(\theta_i^\t-\theta_j^\t)+\frac{\Delta}{2}\cos(\theta_i^\t-\theta_i^{\t+1})+\epsilon J_2\cos 2(\theta_i^\t-\theta_j^\t)+\frac{1}{4}\left(\frac{1}{\epsilon U}-\frac{\Delta}{2}\right)\cos2(\theta_i^\t-\theta_i^{\t+1})\right).\nonumber
\ee
Equating the above equation with the classical partition function we find
\be
\epsilon J_1=\frac{\Delta}{2}=\tanh^{-1}e^{-2\epsilon\Gamma},\quad {\rm and}~\epsilon J_2=\frac{1}{4}\left(\frac{1}{\epsilon U}-\frac{\Delta}{2}\right)=\frac{1}{4}\left(\frac{1}{\epsilon U}-\epsilon J_1\right),
\ee
Hence the quantum Hamiltonian reads
\be
\epsilon \hat{H}=\sum_{\left< ij\right>}-\epsilon J_1\cos(\theta_i-\theta_j)-\epsilon J_2\cos 2(\theta_i-\theta_j)+\sum_i\frac{1}{2(\epsilon J_1+4\epsilon J_2)}n_i^2-\frac{\ln\coth\epsilon J_1}{2}\cos\pi n_i.
\ee
Redefine $\epsilon J_1\mapsto J_1$, $\epsilon J_2\mapsto J_2$ and $\epsilon \hat{H}\mapsto  \hat{H}$ we obtain
\be
 \hat{H}=\sum_{\left< ij\right>}-J_1\cos(\theta_i-\theta_j)-J_2\cos 2(\theta_i-\theta_j)+\sum_i\frac{1}{2(J_1+4J_2)}n_i^2-\frac{\ln\coth J_1}{2}\cos\pi n_i.\label{eq:hq}
\ee
In the limit of $J_2\ra+\infty$ we arrive at 
\be&&\lim_{J_2\ra+\infty}\hat{H}= -J_1\sum_{\left<ij\right>}\cos(\theta_i-\theta_j)-\frac{\ln\coth J_1}{2}\sum_i\cos\pi n_i.\nn
&&\text{Hilbert space constraint:~} \theta_i-\theta_j=k_{ij}\pi, k_{ij}\in\mathbb{Z}. 
\label{j2divv}\ee which is given in the main text. 

\end{document}